\documentclass{ws-ijmpe}

\begin{document}

\markboth{Aneta Iordanova, STAR Collaboration}{System Size Dependence of Freeze-out Properties at RHIC}

\catchline{}{}{}{}{}

\title{SYSTEM SIZE DEPENDENCE\\ OF FREEZE-OUT PROPERTIES AT RHIC}

\author{\footnotesize ANETA IORDANOVA (for the STAR Collaboration)}

\address{Physics Department, University of Illinois at Chicago, 845 W. Taylor\\
Chicago, Illinois, 60607, USA\\
aiorda1@uic.edu}

\author{OLGA BARANNIKOVA}

\address{Physics Department, University of Illinois at Chicago, 845 W. Taylor\\
Chicago, Illinois, 60607, USA\\
barannik@uic.edu}

\author{RICHARD HOLLIS}

\address{Physics Department, University of Illinois at Chicago, 845 W. Taylor\\
Chicago, Illinois, 60607, USA\\
rholli3@uic.edu}

\maketitle

\begin{history}
\received{(received date)}
\revised{(revised date)}
\end{history}

\begin{abstract}
The STAR experiment at RHIC has measured identified $\pi^{\pm}$, $K^{\pm}$
and $p(\overline{p})$ spectra and ratios from $\sqrt{s_{NN}} = 62.4$
and 200~GeV Cu+Cu collisions.  The new Cu+Cu results are studied with
hydro-motivated blast-wave and statistical model frameworks in order to
characterize the freeze-out properties of this system.
Along with measurements from Au+Au and p+p collisions, the
obtained freeze-out parameters are discussed  as a function of
collision energy, system size, centrality and inferred energy density.
This multi-dimensional systematic study reveals the importance of the
collision geometry and furthers our understanding of the QCD phases.
\end{abstract}

\section{Introduction}

Identified $\pi^{\pm}$, $K^{\pm}$ and $p(\overline{p})$  particle spectra
in heavy-ion collisions at different center-of-mass energies provide a
unique tool to explore the QCD phase diagram.~\cite{cite:QCD_Diagram}
Analysis of Cu+Cu data,
collected by the STAR experiment, extends the systematic studies of
bulk properties by addressing the energy and system size dependence of
freeze-out parameters at RHIC. 


Previous studies of the freeze-out parameters in 200 and 62.4 GeV Au+Au
collisions within a chemical and kinetic equilibrium model, showed a
similar chemical freeze-out temperature and an increasing radial
flow with centrality.~\cite{200spectra}~\cite{62spectra}
The centrality independence of the extracted chemical freeze-out
temperature indicates that, for different initial conditions, collisions
evolve to the same freeze-out condition. The values of the chemical 
freeze-out temperature and the critical temperature, predicted by the 
Lattice QCD, are close for all centralities, which suggests that the 
chemical freeze-out
coincides with hadronization and therefore provides a lower limit estimate 
for a temperature of prehadronic state.~\cite{Olgaposter} 
For kinetic freeze-out, which occurs
later, an increase in centrality results in a decreasing temperature
and increasing flow velocity.  Most measured bulk properties 
show a smooth systematic change with the charged hadron multiplicity
and appear to form smooth curves with similar results from lower-energy
collisions. 

With systematic studies of yields and momentum spectra in Cu+Cu collisions,
at the same center-of-mass energies as for Au+Au collisions, we constrain
the initial conditions of the created medium and address the degree
of its thermalization.

\section{Identified Particle Spectra}

For this analysis, charged particles are identified using their
ionization energy loss $dE/dx$ in the STAR Time Projection Chamber
(TPC)~\cite{STAR_TPC}, which is situated in a uniform 0.5~T magnetic
field along the beam line.  The Cu+Cu events are divided into centrality
classes based on the multiplicity distribution within $|\eta|<0.5$.
Six 10\% bins are used to represent the top 60\% of the inelastic
collision cross-section.

The raw (uncorrected) particle yields are obtained from the mean
$\langle dE/dx \rangle$ distributions.  Unidentified spectra are first
divided into bins of transverse momentum ($\Delta p_{T}=0.25$MeV),
one bin in rapidity ($|y|<0.1$), and the six bins in centrality.
Next, projections of $dE/dx$ for a given momentum and centrality bin
are fit with a four-Gausian function, representing the different
particle species ($\pi$, $K$, $p$ and $e$) which are clearly separated
in this transverse momentum region $0.2<p_{T}<0.8$~GeV/c for $\pi^{\pm}$,
$k^{\pm}$ and $0.4<p_{T}<1.2$~GeV/c for $p(\overline{p}$,);
the integral of each Gaussian provides the yield.
Once identified, these yields are further corrected for
detector acceptance, tracking inefficiency and background contributions.
The described analysis technique~\cite{200spectra} is consistent
for all low $p_{T}$ measurements for different center-of-mass energies and 
colliding systems.

The particle spectra for $\pi^{\pm}$, $K^{\pm}$ and $p(\overline{p})$  in 
Cu+Cu collisions are measured for both $\sqrt{s_{NN}} = 62.4$ and 200~GeV. 
The particle and anti-particle spectral shapes are similar for each
centrality class.  For both center-of-mass energies, a mass dependence
is observed in the slope of the particle spectra.

\section{Kinetic Freeze-out Properties}

The hydro-motivated Blast-wave model assumes a boosted thermal source
in transverse and longitudinal directions.\cite{BlastWaveModel}
From simultaneous fits to the $\pi^{\pm}$, $K^{\pm}$ and 
$p(\overline{p})$ spectra in each centrality class three fit parameters
are extracted: the flow velocity ($\beta$), the kinetic freeze-out
temperature ($T_{kin}$) and the shape of the flow profile ($n$). 
For the case of pions, the range below 0.5~GeV/c is excluded from the
fits in order to reduce effects from resonance contributions.

\begin{table}[h]
\centering
\begin{tabular}{|l|c|c|c|c|}
\hline
 & \multicolumn{2}{c|}{200 GeV} & \multicolumn{2}{c|}{62.4 GeV}\\
 Parameter & Cu+Cu           & Au+Au       & Cu+Cu       & Au+Au \\ \hline
 $T_{kin} (MeV)$ & 112$\pm$9 & 89$\pm$12   & 115$\pm$7     & 93.8$\pm$4.1  
                                                                   \\ \hline
 $\beta$   & 0.52$\pm$0.07 & 0.59$\pm$0.06 & 0.48$\pm$0.06 & 0.54$\pm$0.01
                                                                   \\ \hline
\end{tabular}
\caption{\label{KineticTable}
The obtained preliminary kinetic freeze-out fit parameters for 0-10\% central
Cu+Cu collisions at 62.4 and 200~GeV and the corresponding fit parameters
for 0-5\% central Au+Au collisions.}
\end{table}

\begin{figure}[h]
\begin{minipage}{15pc}
\centerline{\psfig{file=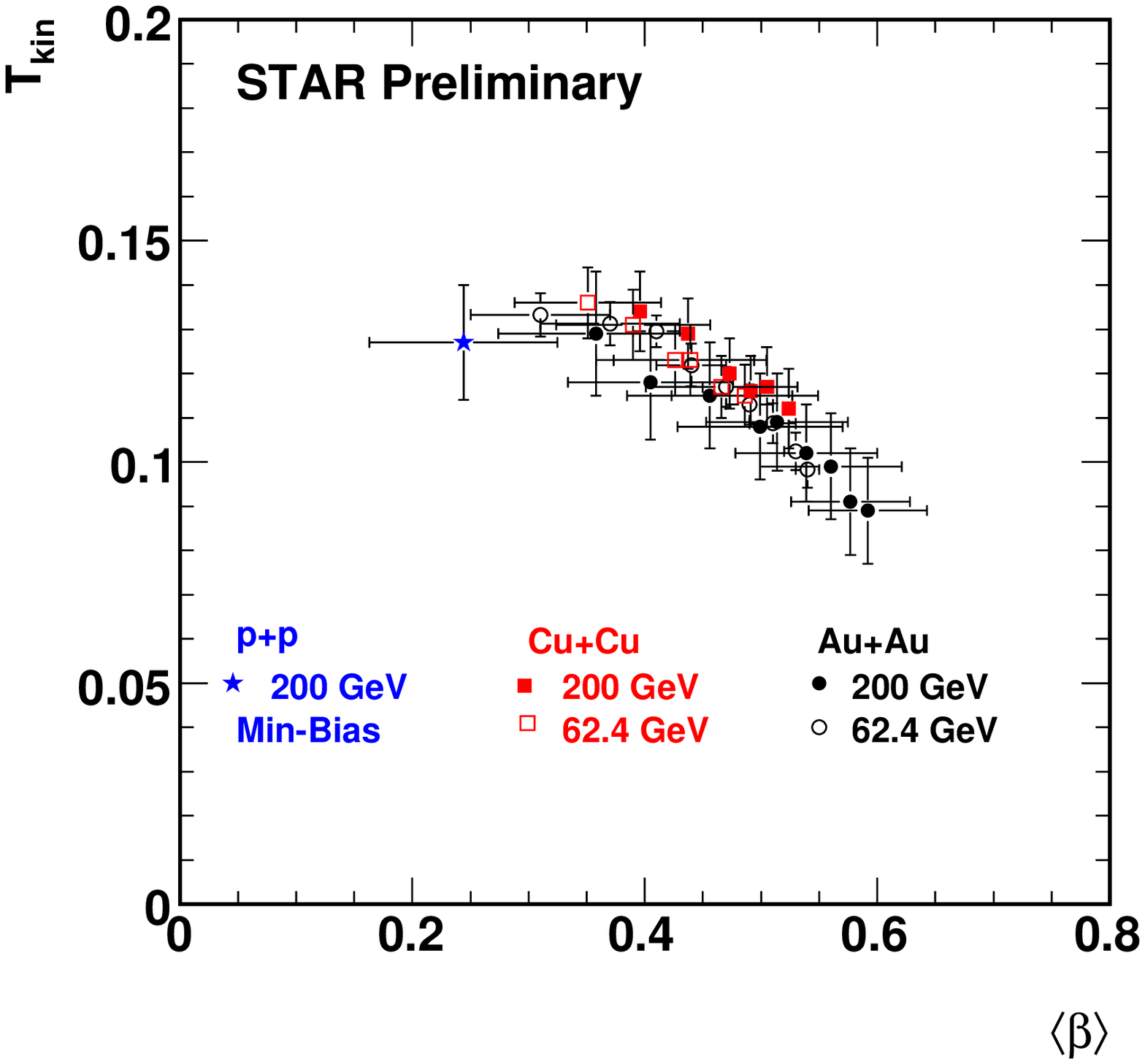, width=1.00\textwidth}}
\end{minipage}
\hspace{-1pc}
\begin{minipage}{15pc}
\centerline{\psfig{file=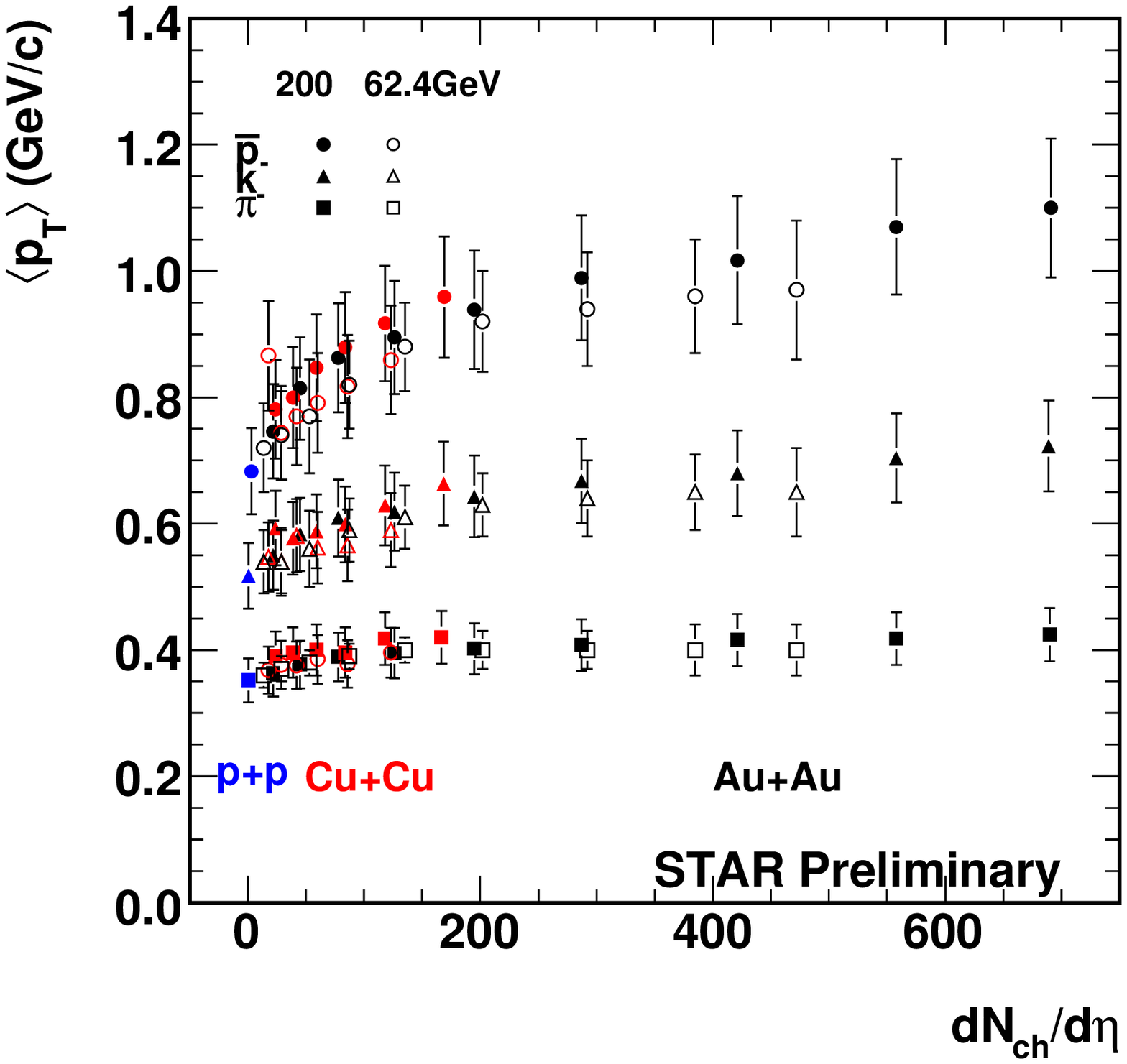, width=1.00\textwidth}}
\end{minipage}
\vspace*{8pt}
\caption{
\label{Tkinvsbeta}
Left panel: Kinetic freeze-out temperature $T_{kin}$ versus flow
velocity $\beta$ for 62.4 and 200~GeV Au+Au (black) and Cu+Cu (red)
collisions. For comparison,
results for minimum bias p+p collisions at 200~GeV are also shown (blue). 
Right panel: Integrated anti-particle $\langle p_{T} \rangle$ for Au+Au
(black) and Cu+Cu (red) collisions as a function of charged hadron
multiplicity density ($dN_{ch}/d\eta$) at mid-rapidity for 62.4 and
200~GeV. Minimum bias p+p collisions at 200~GeV are also shown (blue).}
\end{figure}

\begin{figure}[th]
\begin{minipage}{15pc}
\centerline{\psfig{file=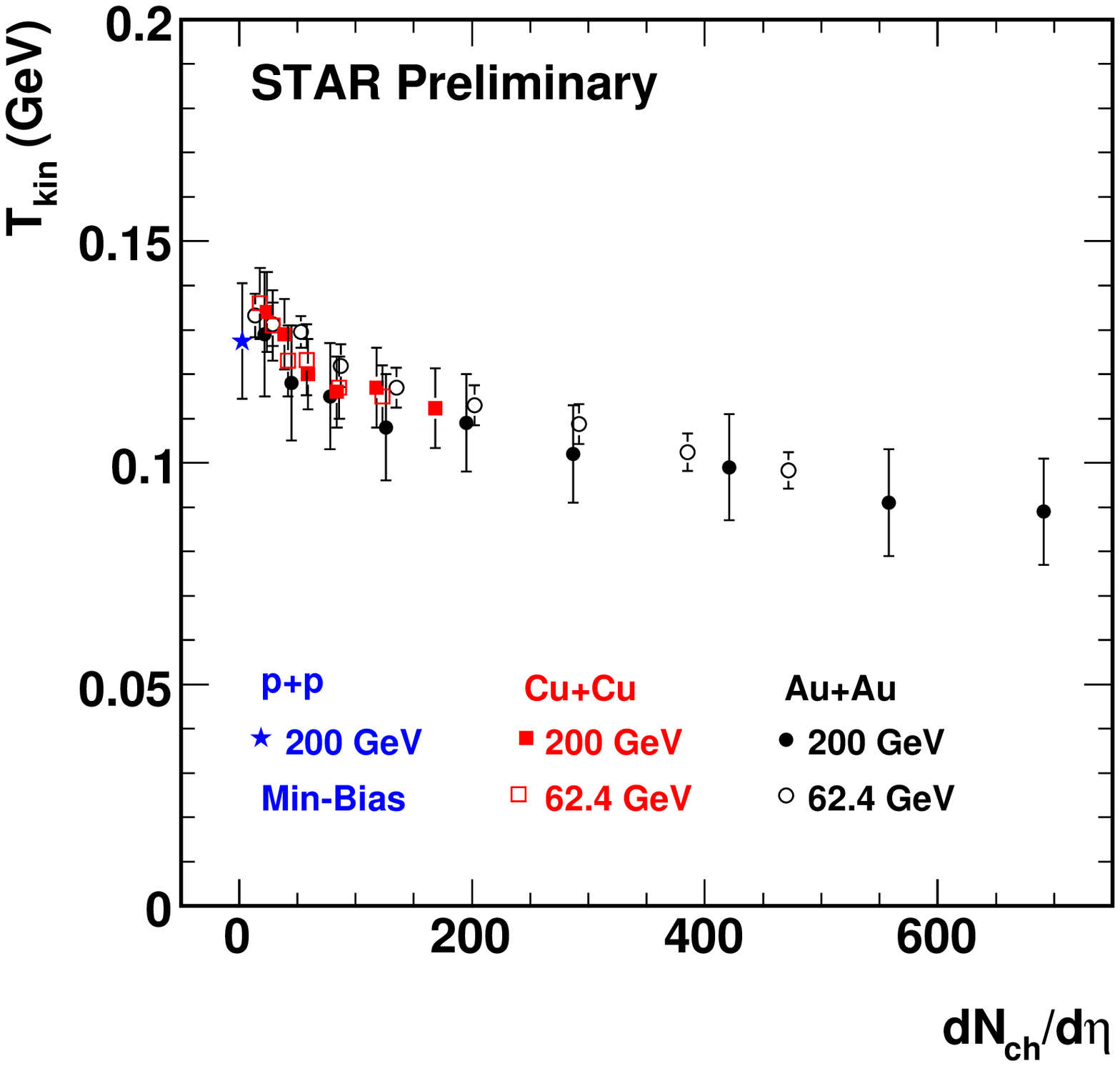, width=1.00\textwidth}}
\end{minipage}
\hspace{-1pc}
\begin{minipage}{15pc}
\centerline{\psfig{file=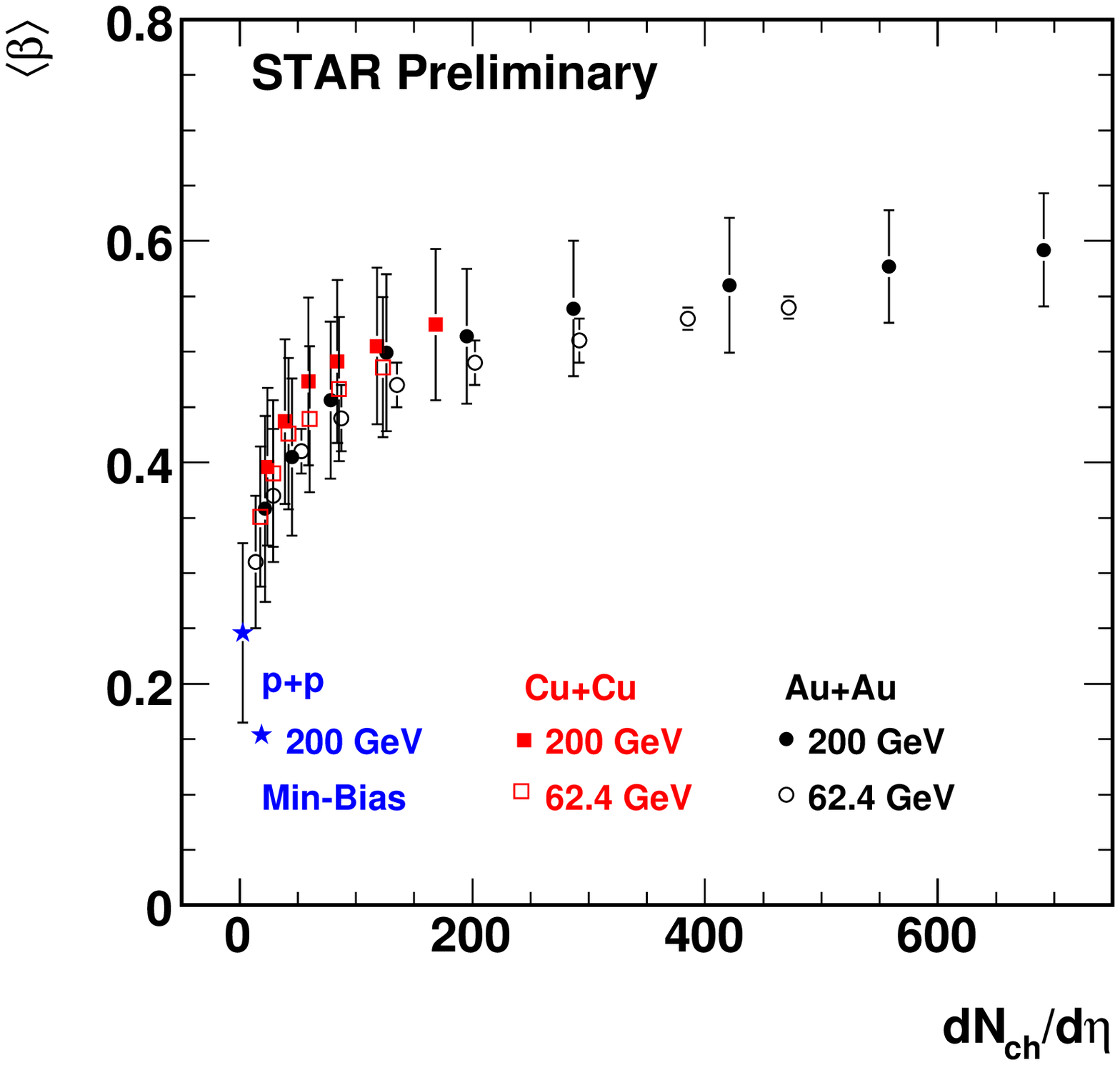, width=1.00\textwidth}}
\end{minipage}
\vspace*{8pt}
\caption{
  \label{Tkin_Nch}
Systematics of kinetic freeze-out parameters for 62.4 and 200~GeV Au+Au
(black) and Cu+Cu (red) collisions as a function of charged hadron
multiplicity density $dN_{ch}/d\eta$ at mid-rapidity.  For comparison,
results for minimum bias p+p collisions at 200~GeV are also shown (blue).}
\end{figure}

Results of the kinetic fit parameters $T_{kin}$ and $\beta$ for Cu+Cu
and Au+Au collisions are shown in Fig.~\ref{Tkinvsbeta} and
Table~\ref{KineticTable}\footnote{Note that we are not interpreting the 
extracted small, but finite, value of $\beta$ in p+p collisions as an 
indication of collectivity in elementary systems. 
The observed value is a reflection of the well known difference in 
the $m_{T}$ spectral shapes of $\pi^{\pm}$, $K^{\pm}$ and $p(\overline{p})$  
in p+p and A+A collisions.~\cite{200spectra}}.  
The studies show that all particle spectra
for $\pi^{\pm}$, $K^{\pm}$ and $p(\overline{p})$ in Cu+Cu and Au+Au
systems can be described by a common set of freeze-out parameters.
Furthermore, the freeze-out parameters extracted are similar for an 
equivalent number of produced charged-particles, $N_{ch}$,
independent of both collision system and center-of-mass energy. 
As can be seen from Fig.~\ref{Tkin_Nch}, the centrality dependence of 
$T_{kin}$ and $\beta$ evolves smoothly from p+p to Cu+Cu and Au+Au 
collisions.

The particle mean-$p_{T}$ are found to increase with centrality,
depicted as $N_{ch}$ in Fig.~\ref{Tkinvsbeta} (left panel).  As for the
kinetic freeze-out parameters, a smooth centrality dependence of the
particle mean-$p_{T}$, for all particle species, is observed
from p+p to Cu+Cu and Au+Au collisions.  The mean-$p_{T}$ results are
produced from Blast-wave fits to the $K^{\pm}$ and $p(\overline{p})$ 
spectra and from Bose-Einstein fits over the entire fiducial spectra
for $\pi^{\pm}$.  Such remarkable similarities in the obtained kinetic
freeze-out parameters as a function of produced charged-particles
observed for both Cu+Cu and Au+Au systems and for both center-of-mass
energies, 200 and 62.4~GeV, raises the question: {\it Are the bulk properties
entropy driven?}

\section{Testing Hadro-Chemistry}

In the statistical model framework, particle yield ratios can be used to 
provide information on the chemical freeze-out properties of the system,
including the chemical temperature at chemical freeze-out, strangeness 
and baryon production.~\cite{StatModel} 

\begin{figure}[th]
\begin{minipage}{15pc}
\centerline{\psfig{file=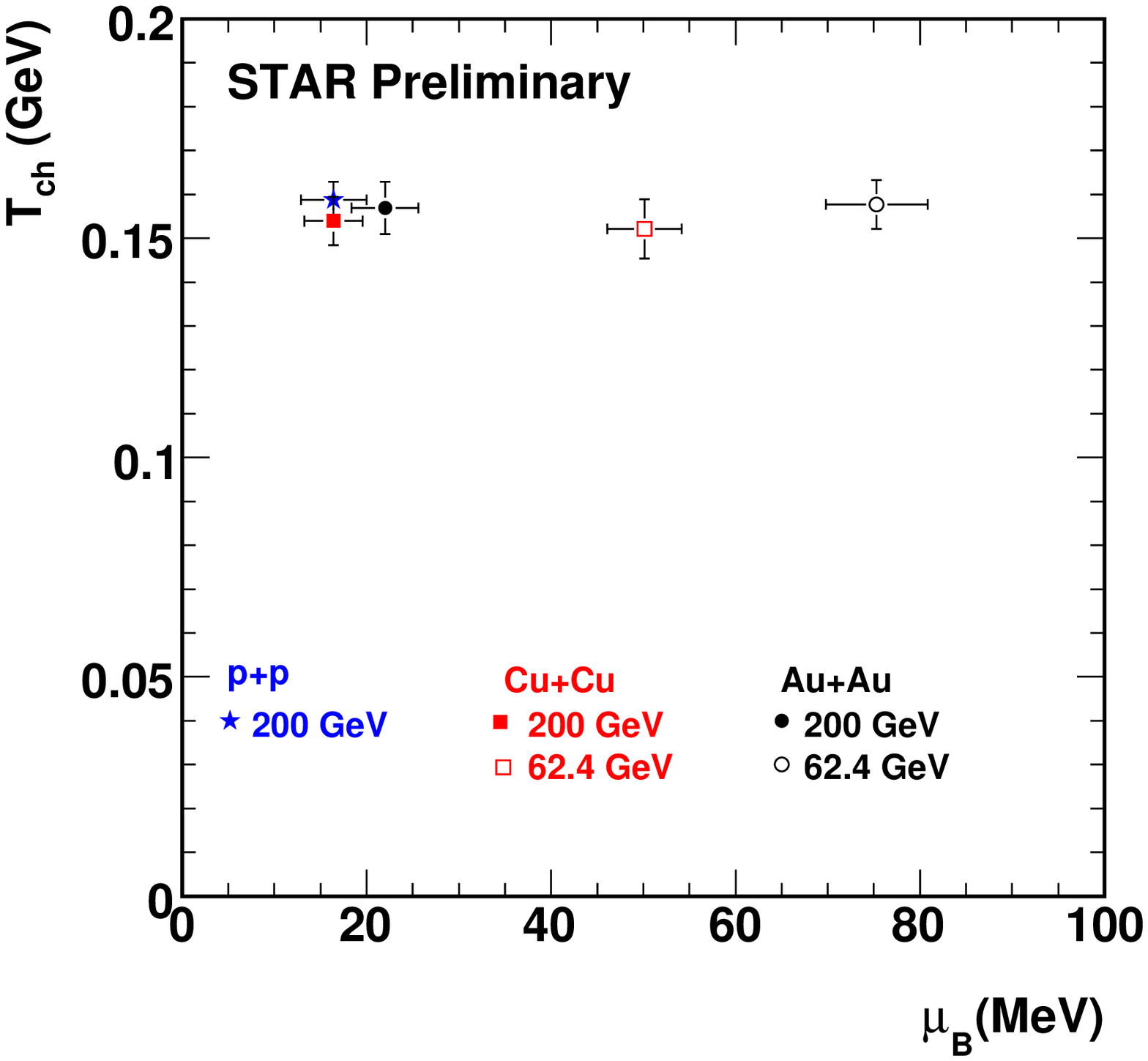, width=1.00\textwidth}}
\end{minipage}
\hspace{-1pc}
\begin{minipage}{15pc}
\centerline{\psfig{file=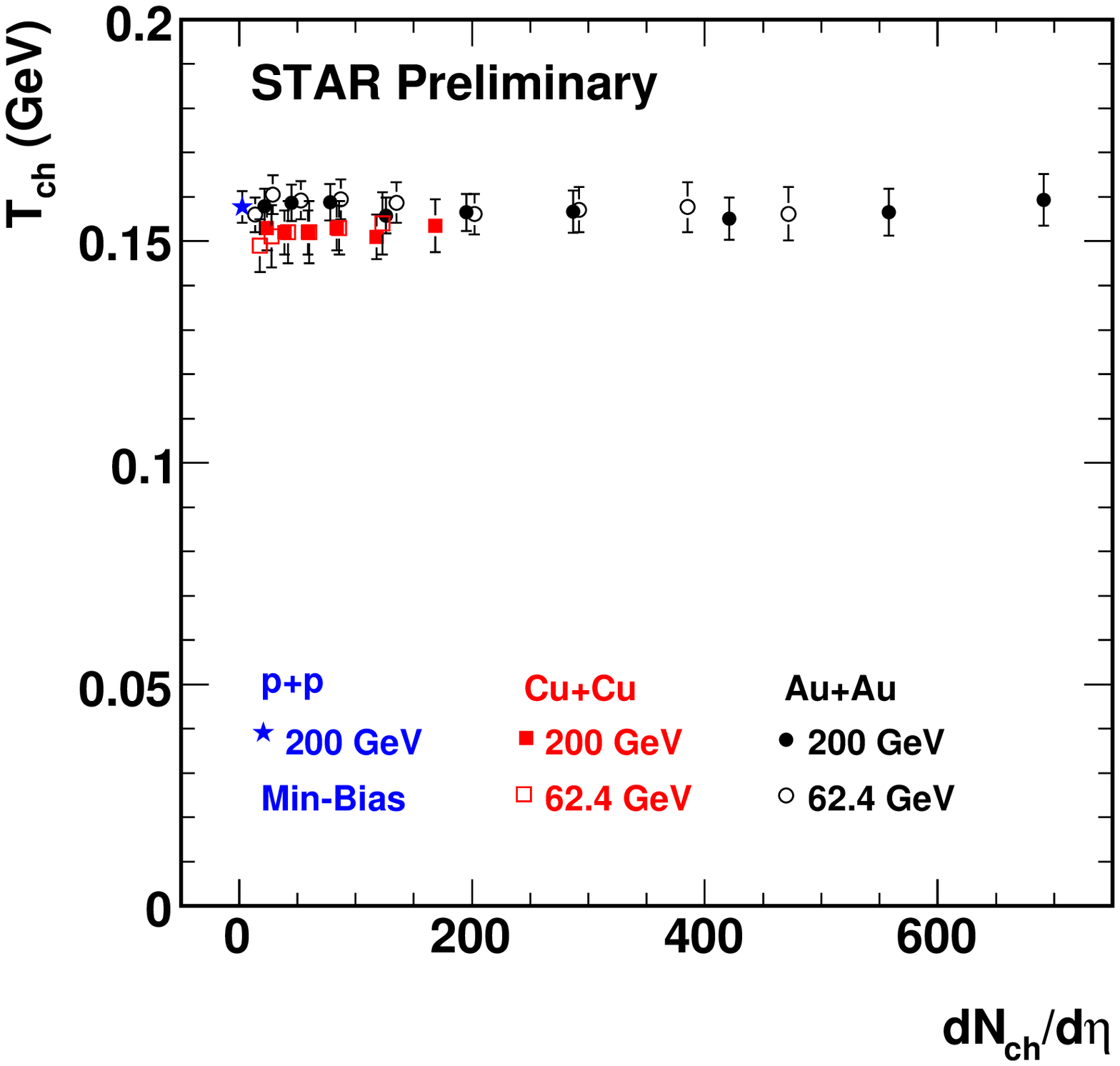, width=1.00\textwidth}}
\end{minipage}
\vspace*{8pt}
\caption{\label{Tch_Nch}
Left panel: Chemical freeze-out temperature $T_{ch}$ versus baryon chemical
potential $\mu_{B}$ for 0-5\% central Au+Au (black) and 0-10\% Cu+Cu (red)
collisions at 62.4 and 200~GeV. For comparison, results for minimum bias
p+p collisions at 200~GeV are also shown (blue).
Right panel: Systematics of chemical freeze-out temperature $T_{ch}$
for 62.4 and 200~GeV Au+Au (black) and Cu+Cu (red) collisions as a function
of charged hadron multiplicity density $dN_{ch}/d\eta$ at mid-rapidity.
Results for minimum bias p+p collisions at 200~GeV are also shown (blue).}
\end{figure}

The Cu+Cu particle ratios for each centrality bin are fit to derive
four fit parameters: 
the chemical freeze-out temperature ($T_{ch}$), 
the baryon chemical potential ($\mu_{B}$), 
the strangeness chemical potential ($\mu_{S}$)
and the strangeness suppression ($\gamma_{S}$).

Fig.~\ref{Tch_Nch} summarizes the systematics of chemical freeze-out
parameters $T_{ch}$ and $\mu_{B}$ for 62.4 and 200~Cu+Cu collisions,
compared to Au+Au and p+p results.  The left panel shows that a universal
chemical freeze-out temperature is observed for all studied systems.
The chemical freeze-out temperature for different colliding systems
shows the same systematic dependence as a function of charged particle 
multiplicity density at mid-rapidity, as illustrated in the right panel 
of Fig.~\ref{Tch_Nch}.
The apparent absence of a decreasing trend in $T_{ch}$, in contrast with the  
$T_{kin}$ systematics, suggests that collisions with different initial 
energy densities evolve to the same  chemical freeze-out, which then must 
coincide with hadronization.

\section{Summary}

The STAR collaboration has enlarged the variety of hadron spectra
measurements at RHIC by providing new results for Cu+Cu collisions at
two different center-of-mass energies, $\sqrt{s_{NN}} = 62.4$ and 200~GeV.
The data have been studied within the framework of both a blast-wave and
a statistical model to investigate the final hadronic state properties
as a function of collision energy, system size, centrality and inferred
energy density.

This multi-dimensional systematic study of the kinetic and chemical
freeze-out parameters has revealed remarkable similarities in the
description of the studied systems. The freeze-out parameters are found
to be independent of the collision system and center-of mass energy.
Furthermore, the parameters from different systems show a smooth evolution 
with centrality and similar properties at the same number of produced
charged-particles, which is believed to be connected to the initial energy 
density or the size of the initial collision geometric overlap.

The bulk properties are most probably determined in the initial stages
of the collision and driven by the initial energy density.

\section*{Acknowledgements}
We thank the RHIC Operations Group and RCF at BNL, and the
NERSC Center at LBNL for their support. This work was supported
in part by the Offices of NP and HEP within the U.S. DOE Office 
of Science; the U.S. NSF; the BMBF of Germany; CNRS/IN2P3, RA, RPL, and
EMN of France; EPSRC of the United Kingdom; FAPESP of Brazil;
the Russian Ministry of Science and Technology; the Ministry of
Education and the NNSFC of China; IRP and GA of the Czech Republic,
FOM of the Netherlands, DAE, DST, and CSIR of the Government
of India; Swiss NSF; the Polish State Committee for Scientific 
Research; SRDA of Slovakia, and the Korea Sci. \& Eng. Foundation.


\begin{thebibliography}{0}

\bibitem{cite:QCD_Diagram} F.Karsch, {\it J.Phys.Conf.Ser.}  {\bf 46} (2006) 122.

\bibitem{200spectra} J.Adams et al., {\it Phys. Rev. Lett.} {\bf 92} (2004) 112301.

\bibitem{62spectra} L.Molnar et al., {\it arXiv:} nucl-ex/0507027.

\bibitem{Olgaposter} O.Barannikova et al., {\it arXiv:} nucl-ex/0403014. 

\bibitem{BlastWaveModel} E. Schnedermann, J. Sollfrank and U. Heinz, {\it Phys. Rev.} {\bf C48} (1993) 2462.

\bibitem{StatModel} P. Braun-Munzinger, I. Heppe and J. Stachel, {\it Phys. Lett.} {\bf B465} (1999) 15.

\bibitem{STAR_TPC} M. Anderson et al., {\it Nucl. Instrum. Meth.} {\bf A499} (2003) 659.

\end{thebibliography}
\end{document}